\documentclass[10pt]{wlscirep}
\usepackage[utf8]{inputenc}
\usepackage[T1]{fontenc}
\usepackage{bm}

\newcommand{\figref}[1]{Fig.~\ref{#1}}

\newcommand{\secref}[1]{``{#1}''~section}
\renewcommand{\eqref}[1]{~(\ref{#1})}

\newcommand{\x}{\bm{x}}
\newcommand{\lnp}[2]{\mathrm{ln}p_{\bm{#1}}(#2)}
\newcommand{\pz}{p_{\bm{}{\theta}}(\bm{z})}
\newcommand{\pzlx}{p_{\bm{\theta}}(\bm{z}|\x^{(i)})}
\newcommand{\pxlz}{p_{\bm{\theta}}(\x^{(i)}|\bm{z})}
\newcommand{\qzlx}{q_{\bm{\phi}}(\bm{z}|\x^{(i)})}
\newcommand{\intz}[1]{\int {#1} d\bm{z}}
\newcommand{\vlb}{\mathcal{L}(\x^{(i)}, \bm{\theta}, \bm{\phi})}
\newif\ifDraft
\Drafttrue

\definecolor{DarkGreen}{rgb}{0.0,0.45,0}
\definecolor{Sakujo}{rgb}{0.9,0.5,0.9}
\definecolor{Grey}{rgb}{0.7,0.7,0.7}
\ifDraft

\else

\fi

\title{Unsupervised Learning Architecture for Classifying the Transient Noise of Interferometric Gravitational-wave Detectors}

\author[1,+]{Yusuke~Sakai}
\author[2, 3]{Yousuke~Itoh}
\author[4]{Piljong~Jung}
\author[5]{Keiko~Kokeyama}
\author[6]{Chihiro~Kozakai}
\author[7]{Katsuko~T.~Nakahira}
\author[8]{Shoichi~Oshino}
\author[9, 10, 11]{Yutaka~Shikano}
\author[1, 12, 13, ++]{Hirotaka~Takahashi}
\author[8]{Takashi~Uchiyama}
\author[7]{Gen~Ueshima}
\author[6]{Tatsuki~Washimi}
\author[8]{Takahiro~Yamamoto}
\author[8]{Takaaki~Yokozawa}

\affil[1]{Research Center for Space Science, Advanced Research Laboratories, Tokyo City University, Setagaya-ku, Tokyo 158-0082, Japan}
\affil[2]{Graduate School of Science, Osaka Metropolitan University, Sumiyoshi-ku, Osaka City, Osaka 558-8585, Japan}
\affil[3]{Nambu Yoichiro Institute of Theoretical and Experimental Physics (NITEP),
Osaka Metropolitan University, Sumiyoshi-ku, Osaka City, Osaka 558-8585, Japan}
\affil[4]{National Institute for Mathematical Sciences, Daejeon 34047, Republic of Korea}
\affil[5]{School of Physics and Astronomy, Cardiff University, The Parade, Cardiff, CF24 3AA United Kingdom}
\affil[6]{Gravitational Wave Science Project, Kamioka branch, National Astronomical Observatory of Japan, Hida City, Gifu 506-1205, Japan}
\affil[7]{Department of Information and Management Systems Engineering, Nagaoka University of Technology, Nagaoka, Niigata 940-2188, Japan}
\affil[8]{Institute for Cosmic Ray Research, KAGRA Observatory, The University of Tokyo, Hida City, Gifu 506-1205, Japan}
\affil[9]{Graduate School of Science and Technology, Gunma University, Maebashi, Gunma 371-8510, Japan }
\affil[10]{Institute for Quantum Studies, Chapman University, Orange, CA 92866, USA}
\affil[11]{JST PRESTO, Kawaguchi, Saitama 332-0012, Japan}
\affil[12]{Institute for Cosmic Ray Research, The University of Tokyo, Kashiwa City, Chiba 277-8582, Japan}
\affil[13]{Earthquake Research Institute, The University of Tokyo, Bunkyo-ku, Tokyo 113-0032, Japan}

\affil[+]{g2191402@tcu.ac.jp}
\affil[++]{hirotaka@tcu.ac.jp}

\begin{abstract}
In the data obtained by laser interferometric gravitational wave detectors, transient noise with non-stationary and non-Gaussian features occurs at a high rate. This often results in problems such as detector instability and the hiding and/or imitation of gravitational-wave signals. This transient noise has various characteristics in the time--frequency representation, which is considered to be associated with environmental and instrumental origins. Classification of transient noise can offer clues for exploring its origin and improving the performance of the detector. One approach for accomplishing this is supervised learning. However, in general, supervised learning requires annotation of the training data, and there are issues with ensuring objectivity in the classification and its corresponding new classes. By contrast, unsupervised learning can reduce the annotation work for the training data and ensure objectivity in the classification and its corresponding new classes. In this study, we propose an unsupervised learning architecture for the classification of transient noise that combines a variational autoencoder and invariant information clustering. To evaluate the effectiveness of the proposed architecture, we used the dataset (time--frequency two-dimensional spectrogram images and labels) of the Laser Interferometer Gravitational-wave Observatory (LIGO) first observation run prepared by the Gravity Spy project. The classes provided by our proposed unsupervised learning architecture were consistent with the labels annotated by the Gravity Spy project, which manifests the potential for the existence of unrevealed classes.
\end{abstract}
\begin{document}
\flushbottom
\maketitle
%
%
\thispagestyle{empty}


\newcounter{intro}
Gravitational waves are distortions of the space--time continuum that propagate (with high probability) at the speed of light. They are emitted during events such as the coalescence of compact star binaries and supernova  explosions. The first observation of a gravitational wave, which was from the coalescence of a black hole binary, was achieved by the Laser Interferometer Gravitational-wave Observatory (LIGO)~\cite{ref:aasi2015advanced} located in Livingston, Louisiana and Hanford, Washington in the USA in September 2015~\cite{ref:Abbott2016}. Subsequently, LIGO and Virgo~\cite{ref:acernese2014advanced} in Europe made three international joint observation runs and observed as many as 90 events of gravitational waves emitted by the coalescence of compact binaries~\cite{ref:Abbott2019,ref:Abbott2020,ref:Abbott2021_a,ref:Abbott2021_b}. Moreover, GEO600~\cite{ref:grote2008status}, in Germany and  KAGRA~\cite{ref:akutsu2018kagra,ref:2021PTEP.2021eA101A,ref:2020arXiv200802921K,ref:2021PTEP.2021eA102A}, in Japan, made a 2-week observation run (O3GK) in April 2020~\cite{ref:2022bkagra,ref:2022akagra}. The subsequent fourth observation run (O4) is planned to be conducted jointly with LIGO, Virgo, and KAGRA.

When searching for a gravitational wave signal in the data from the interferometers, suitable techniques for separating the gravitational waves from instrumental noise in the observed data are essential because the signals of the gravitational waves are generally smaller than the detector noise. The gravitational-wave detector is sensitive to environmental and instrumental states (such as ground motions, air pressure, optics suspensions, fluctuations in the laser, vacuum, and mirror). Consequently, non-stationary and non-Gaussian noise, called ``transient noise'', frequently appears in the detector. Transient noise causes instability in the detector and the hiding and/or imitating of the gravitational-wave signals. The LIGO and Virgo collaboration reported that transient noise with a signal-to-noise ratio $> 6.5$ occurred at a rate of $1.10$ events per minute at LIGO Livingston (LLO) in the first half of the third observation run (O3a) between 1 April 2019, 15:00 UTC and 1 October 2019, 15:00 UTC~\cite{ref:Abbott2020}, and at a rate of $1.17$ events per minute at LLO in the second half of O3 (O3b) between 1 November 2019, 15:00 UTC and 27 March 2020, 17:00 UTC~\cite{ref:Abbott2021_b}, respectively.

Transient noise has various time--frequency characteristics that are related to its causes in the detector. Classifying transient noise could provide us with clues to explore its origins and improve the performance of the detector. Among others, the Gravity Spy project~\cite{ref:Zevin2017,ref:Bahaadini2018,ref:Soni2021,ref:Bahaadini2017} is one such effort to classify transient noise. The Gravity Spy project used the Omicron software~\cite{ref:Robinet2020} to identify the signal of transient noise observed in the time-series data. Thereafter, Omega Scan~\cite{ref:Chatterji2004} was used to create a time--frequency spectrogram around the identified transient noise as two-dimensional (2D) images. Based on a part of these created 2D images, using cloud resources in collaboration with LIGO detector characterisation experts and volunteer citizen scientists for the analysis, 22 types of labels associated with the characteristics or causes of transient noise were annotated. Both images and labels were recorded. Finally, they classified the transient noise in the remaining images by supervised learning using the pre-classified images and labels. As this process shows, the data annotation for machine learning is highly labour-intensive.

Previous studies~\cite{ref:Bini2020unsupervised} using unsupervised classification grouped together similar transient noise in the Gravity Spy dataset~\cite{ref:Bahaadini2018}. Bahaadini et al. used the DIRECT method~\cite{ref:bahaadini2018direct} to analyse the feature embedding learned from the Gravity Spy dataset~\cite{ref:Bahaadini2018} and observed a different class of transient noise from the existing classes. Unsupervised clustering applying transfer learning~\cite{ref:george2018} exhibited a new class of transient noise in addition to the 22 classes of the Gravity Spy project. Moreover, supervised classification using the latest observation O3 dataset presented a new class of transient noise~\cite{ref:Soni2021}.

As unsupervised learning does not require any pre-assigned labels for the training dataset, this architecture is expected to reduce annotation work for the training data, increase the objectivity of the classification, and even classify a new class of the transient noise. Unsupervised learning is also useful in various fields, such as text categorisation, feature representation, and clustering~\cite{ref:Shiping2019,ref:Wenzhong2020,ref:Jinyu2021,ref:Jinyu2022}. In this study, we focus on unsupervised learning using a deep convolutional neural network (CNN) and propose a classification architecture for transient noise. Our proposed architecture consists of two processes: feature learning and classification. In the feature learning process, the features of transient noise are extracted from the time--frequency spectrogram images (2D images) using a variational autoencoder (VAE)~\cite{ref:Kingma2014,ref:Kingma2019}.
In the classification process, invariant information clustering (IIC)~\cite{ref:Ji2019} is used to classify images of the transient noise using features extracted by the encoder of the pre-learned VAE. We applied the proposed architecture to the dataset~\cite{ref:Bahaadini2018} created by the Gravity Spy project of the LIGO observation run 1 (O1)~\cite{ref:Abbott2019} as our input images, examined the validity of the unsupervised classification result, and analysed the correspondence with the labels of the Gravity Spy project.

\section*{Results} \refstepcounter{intro} \label{sec:results}
The result section consists of two subsections: the results of the training process and evaluation of the unsupervised learning architecture.
The Gravity Spy dataset of LIGO O1, which was developed by the Gravity Spy project shown in \figref{fig:data-classes}, was used for training in our proposed architecture.
This dataset contains a total of 8535 transient noises in four time durations: 0.5, 1.0, 2.0, and 4.0 s. Each data unit has a label with one of the 22 types which are related to the origins or characteristics of the transient noise.
The labels annotated by the Gravity Spy project under Zooniverse, which is the online citizen science platform, were used only when evaluating the training results of the proposed architecture.
In addition, the pre-processing of the dataset is shown in \secref{Pre-processing}.

\subsection*{Training process of our architecture}
We investigated the training parameters to use for the VAE as follows. The dimensions of the feature variable $\bm{z}$ were $64, 128, 256, 512$, and $1024$; the training size rate was in the range of $[0.6, 0.9]$ in increments of $0.1$; the learning rate using the Adam\cite{ref:Adam} optimiser with parameters $\beta_1 = 0.9$,  $\beta_2 = 0.999$ (coefficients used for computing running averages of gradient and its square) and $\epsilon=10^{-8}$ (term added to the denominator to improve numerical stability) was in the range of $[5\times 10^{-7}, 5 \times 10^{-2}]$ in increments of one digit; the minibatch size was in the range of $[32, 128]$ in increments of $32$. The maximisation of the lower bound\eqref{eq:vlb} (i.e. let $\delta = -\sum_{i}^ {N}\vlb$) was used as a training objective, and the minimisation of $\delta$ was used for training. The value of $\delta$ does not have a significant effect on the dimension of $\bm{z}$ and the training size rate. By contrast, the learning rate and minibatch size are related to the value of $\delta$ and its stability. The representative parameters for training are shown on the left side of \figref{fig:learning_curve}a), and the training curves using these parameters are shown in \figref{fig:learning_curve}b. Considering Case 1 (black line in \figref{fig:learning_curve}b), the learning rate seems too low and $\delta$ does not decrease. Regarding Case 2 (grey line), the result of the training is not stable, showing the fluctuation in the curve, although $\delta$ has decreased compared with Case 1. In Case 3 (blue line), $\delta$ decreases in both the training and evaluation and seems stable after 100 epochs. Considering these results, for the remainder of the study, the parameters of Case 3 were utilised in the proposed architecture. 

Examples of the reconstructed images of the transient noise generated by the decoder of the VAE at 100 epochs are shown in \figref{fig:learning_curve}c. The characteristics of the reconstructed images seem similar to those of the input images. We confirmed a similar tendency for all the other inputs and reconstructed images.
Therefore, the encoder of the VAE at 100 epochs was applied to the IIC for the classification of the transient noise.

Furthermore, the validity of the features by VAE is shown in Supplemental Material ``Feature Visualization of Transient Noise using t-SNE'' section by visualised features $\bm{z}$, which are projected using t-SNE.

After training the VAE, the training parameters of IIC were also investigated using the pre-trained encoder. The output classes were in the range of $[22, 100]$ in increments of $2$; the output over the classes was in range of $[50, 500]$ in increments of $50$; the classifier number was one of $3, 5, 10, 20$; the learning rate of the Adam optimiser with parameters $\beta_1 = 0.9$, $\beta_2 = 0.999$ and $\epsilon=10^{-8}$ was in the range of $[5\times 10^{-7}, 5 \times 10^{-2}]$ in increments of one digit; the minibatch size was in the range of $[64, 256]$ in increments of $32$. Owing to the training, the mutual information from \eqref{eq:maximum_mutual_info} was high, between $30$ and $40$ output classes, which is consistent with the fact that the subclasses are implied in the dataset.
When the output over classes and the classifiers change, the mutual information does not seem to change. In this study, the IIC parameters shown on the left side of \figref{fig:learning_curve}a were used for the classification. In addition, considering the parameters of spectral clustering with multiple classifiers, the number of classifiers $K=5$, and the number of classes $C=36$. These values are the optimal performance for classification using the accuracy shown in \secref{Discussion}.
The training for the VAE and IIC with a 128 mini-batch size took approximately 1.0 h/100 epochs and approximately 0.3 h/100 epochs, respectively, using two NVIDIA GeForce RTX 2080 Ti GPUs, an Intel Xeon CPU E5-2637 v4 (core 8), and 125 GB of main memory.

\subsection*{Evaluation of our architecture}
The evaluation results are presented in this section. The proposed architecture shown in \secref{Proposed Architecture} was trained using the pre-processing dataset described in \secref{Pre-processing}.

\figref{fig:all_class} shows a randomly selected image from each class (representative image) and similar images that have a high degree of similarity to the representative image in a class. These similar images are derived from the cosine similarity~\cite{ref:murphy2012machine} between the representative image and the other images, using an affinity matrix which is calculated by spectral clustering.

The representative images seem to have different characteristics for each class, and similar images are close to their representative images. Moreover, the image of class (15) in \figref{fig:all_class} shows that the classifier recognises the same class even if the data are shifted in the time direction. Therefore, training that does not depend on the perturbation in the time duration is achieved by pre-processing the dataset. 

To investigate the correspondence between the results of supervised and unsupervised learning, a confusion matrix using the Gravity Spy labels is shown in \figref{fig:cm}. Considering the classes (classes (6) (``1080Lines''), (8) (``Repeating\_Blips''), (14) (``Chirp''), (18) (``Helix''), and (24) (``Scratchy'')), unsupervised learning classifies the Gravity Spy labels (noted parentheses) as one class.
In addition, the output images by the classifier shown in \figref{fig:all_class} are similar to those of the Gravity Spy labels. 

The ``Scattered\_Light'' class is separated into classes (2), (3), (11), and (16) on the confusion matrix, respectively. These classes are classified into different classes on unsupervised learning, whereas their characteristics are similar to \figref{fig:all_class}. A previous study~\cite{ref:Soni2021} on supervised learning with the Gravity Spy labels indicated the existence of a subclass that might be in the ``Scattered\_Light'' class.
The unsupervised classification yielded the same results as in the previous study, indicating the existence of a subclass of the ``Scattered\_Light'' class.

Considering the ``Blip'' and ``Koi\_Fish'' classes, both classes are separated into multiple classes as shown in \figref{fig:cm}. The representative images and their similarity images from separating the classes are shown in \figref{fig:similar_sample}, where the similarity images are sorted in descending order and are sampled randomly from the cosine similarity to the representative image. Each separating class is grouped into its own class, even for images with low cosine similarity. 
The images of the classes separated from ``Blip'' have a common Gravity Spy label. Moreover, the frequency growth of the spectrogram image for classes (9), (20), and (30) looks roughly similar, and the unsupervised classification classifies each class using their characteristics details. Similar results can be observed in ``Koi\_Fish'' (class(5) and class(7)). Therefore, the images of ``Blip'' or ``Koi\_Fish'' may be classified into more detailed subclasses.
 
The ``Paired\_Doves'', ``Wandering\_Line'', and ``Air\_Compressor'' classes are a few of the samples in the dataset (\figref{fig:data-classes}b). 
``Air\_Compressor'' is classified into one class; however, the other classes are not classified into any unique classes in the unsupervised classification. We assume that ``Air\_Compressor'' is a class that cannot be divided further. Therefore, it is classified into one class, even with few data. 
Conversely, ``Paired\_Doves'' and ``Wandering\_Line'' are assumed to have more subclasses. The reason why they are not classified into a specific class can be explained by the fact that a limited amount of transient noise is classified into ``Paired\_Doves'' and ``Wandering\_Line''.

The ``None\_of\_the\_Above'' class of the O1 dataset comprises data that do not belong to any other Gravity Spy labels. The unsupervised classification does not classify these data into unique classes; instead, it distributes them into various class types. This result is consistent with a previous study by Bahaadini et al.~\cite{ref:Bahaadini2018}. In fact, Soni et al.~\cite{ref:Soni2021} used the O3 dataset~\cite{ref:Abbott2020} and reported that several of the ``None\_of\_the\_Above'' appear in the ``Blip'' class or the new population of ``Scattering\_Light''. A similar classification result is expected when applying our architecture to the O3 dataset and retraining it.

Based on the above results, the data of the Gravity Spy labels that are classified into multiple classes in unsupervised classification are shown in grey in the ``Estimated number of class'' in \figref{fig:cm}. These data that are separated from the Gravity Spy labels may imply the existence of subclasses.

\section*{Discussion} \refstepcounter{intro} \label{sec:discussion}
Let the number of Gravity Spy classes (labels) be $C^{\prime} = 22$ and the classified result (vector) whose unsupervised class is the $i$-th class be $\bm{v}^{(i)} \in \mathbb{R}^{C^{\prime}}$, where $i \in c = \{0, \dots, 35\}$. Alternatively, $v^{(i)}_j$, indicating the $j$-th component of $\bm{v}^{(i)}$, is the number of the $j$-th images, and the Gravity Spy label is classified as the $i$-th unsupervised class. The total number of classified $i$-th unsupervised classes is expressed by the $L^1$ norm~\cite{ref:murphy2012machine} of $\bm{v}^{(i)}$ (i.e. $|\bm{v}^{(i)}|_1 = \sum_{j=1}^{C^{\prime}}|v^{(i)}_j|$). The $j$-th component of a normalised vector $\bm{v}^{(i)} / |\bm{v}^{(i)}|_1$ is the ratio of the $j$-th image of the Gravity Spy label on the $i$-th unsupervised class. Therefore, we define the accuracy of unsupervised learning as
\begin{equation}
   A = \sum_{i=1}^{C}
  \cfrac{\mathrm{max} (\bm{v}^{(i)})}
  {|\bm{v}^{(i)}|_1}.
  \label{eq:accuracy}
\end{equation}
It should be noted that the confusion matrix shown in \figref{fig:cm} is not a square matrix, and its indices of unsupervised labels (columns) depend on the initial values of training. Therefore it is difficult to define the evaluation indicators, such as recall, precision, and F-measure. The accuracy of the proposed architecture was $90.9\%$, where the total number of unsupervised classes was set to $C=36$. Comparatively, although \eqref{eq:accuracy} is a slightly different definition from the usual definition of the accuracy of supervised learning, the supervised learning of the Gravity Spy project~\cite{ref:Zevin2017} achieved $97.1\%$ accuracy on the testing data using the same dataset as that used here. Furthermore, we compared our results with those (shown in Table~I of reference~\cite{ref:george2018}) of different CNN models, such as Google Inception~\cite{ref:szegedy2015} (with versions 2 and 3), Microsoft ResNet~\cite{ref:kaiming2016}, VGG~\cite{ref:karen2014} (with 16 and 19 layers), and the retrained CNN model based on the Gravity Spy project~\cite{ref:Zevin2017,ref:Bahaadini2017}. Google Inception, ResNet, and VGG are the most popular image recognition architectures, all of which were submitted to the ILSVRC competitions~\cite{ref:jia2009}. Note that all models used the same dataset (Gravity Spy dataset of LIGO O1). The accuracy was more than 96\% for all models. Although the accuracy of our model is less than the that of above models, unsupervised learning has the advantage that data annotations are not required, and our model has the potential to suggest the existence of subclasses, as shown in \secref{Evaluation of our Architecture}.

Let us now examine the classification results in \figref{fig:cm}, one of the factors that decrease the accuracy of unsupervised learning in \eqref{eq:accuracy}.
The representative images of the major characteristics and images of their low similarities are shown in \figref{fig:sample_worse}. Considering classes (0) and (35), the classifier is able to identify the global features of images because the images are similar to the representative images that also exist in the data of other Gravity Spy labels. Regarding classes (13) and (34), the classifier cannot recognise the images properly and may be learning the background features. This problem can be solved by adjusting the neural-network configuration.
Moreover, regarding class (26), it is observed that the minor images (such as ``Power\_Line'') are mixed with the major class (``Air\_Compressor''). The same result can also be observed for class (32). Because the characteristics of both images are similar, it is possible that both noises have similar characteristics. Additionally, a comparison of the classification results shown in \figref{fig:cm} with the feature visualisation using t-SNE is discussed in Supplemental Material ``Feature Visualization of Transient Noise using t-SNE'' section.
Based on the above results, we can confirm the consistency between the label annotated by the Gravity spy project and the class provided by our proposed unsupervised learning architecture and provide the potential for the existence of the unrevealed classes.

Subsequently, we will build a system for the classification of transient noise using the proposed architecture in KAGRA. In addition, we will extend our architecture to self-supervised learning~\cite{ref:lee2013pseudo} to enhance the accuracy of the classification. This algorithm trains the data of a specific label, known as the \textit{golden set}~\cite{ref:Zevin2017}, which generates a \textit{pseudo label} to the given dataset and retrains it.
Using the new classes classified by unsupervised learning, the semi-supervised learning can help reduce the annotation process for the training and can solve the problem of ensuring objectivity in the classification. We would like to construct a semi-supervised architecture that incorporates the advantages of both Gravity Spy's supervised and unsupervised learning.

\section*{Method} \refstepcounter{intro} \label{sec:method}
The proposed unsupervised learning method consists of two architectures: a variational autoencoder (VAE) and invariant information clustering (IIC). The VAE is used to learn the features from the time--frequency spectrogram (2D images) of transient noise, and the IIC classifies the transient noise from the features that are learned by the encoder of the VAE. Before we present the details of the method, we explain the target dataset.

\subsection*{Target dataset}
\begin{figure}[t]
  \centering
  \includegraphics[scale=1.8]{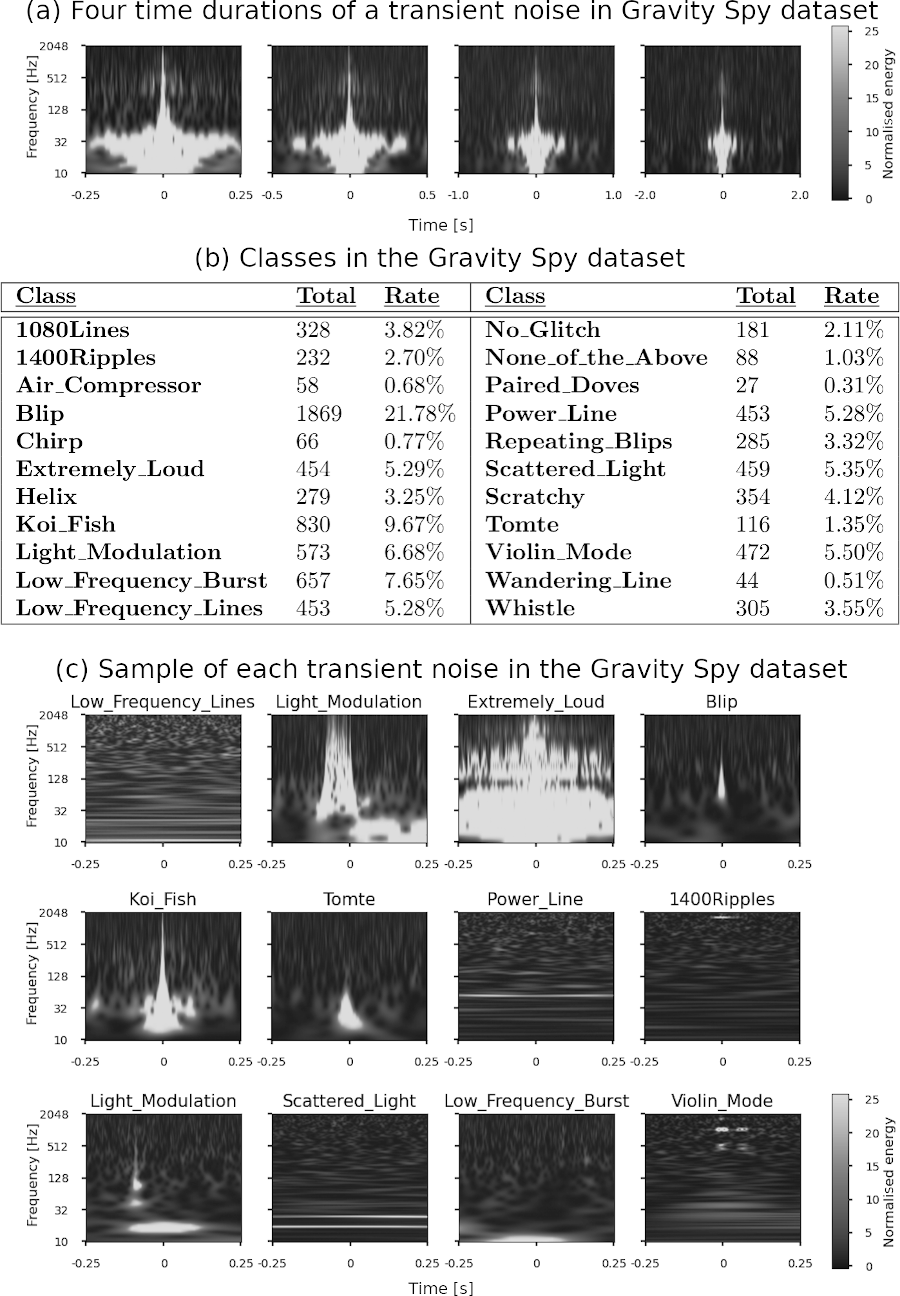}
  \caption{
    (a) Example of 2D image of the time--frequency spectrogram of transient noise in the Gravity Spy dataset. Regarding each transient noise, four time durations (0.5, 1.0, 2.0, and 4.0 s from the left of the figure) are recorded from the centre time.
    (b) Table showing all the classes, the number of data, and its ratio to the number in the Gravity Spy dataset. There are 22 classes in total, and each of 21 classes is given a name related to an occurrence cause or a characteristic of the shape on the spectrogram of transient noise. The other is ``None\_of\_the\_Above'', which does not belong to any class.
    (c) Example of the image for each class in the Gravity Spy dataset.
    The figure shows 12 of the 22 classes of the transient noise with 0.5 s.
}\label{fig:data-classes}
\end{figure}

\begin{figure}[t]
  \centering
  \includegraphics[scale=1.8]{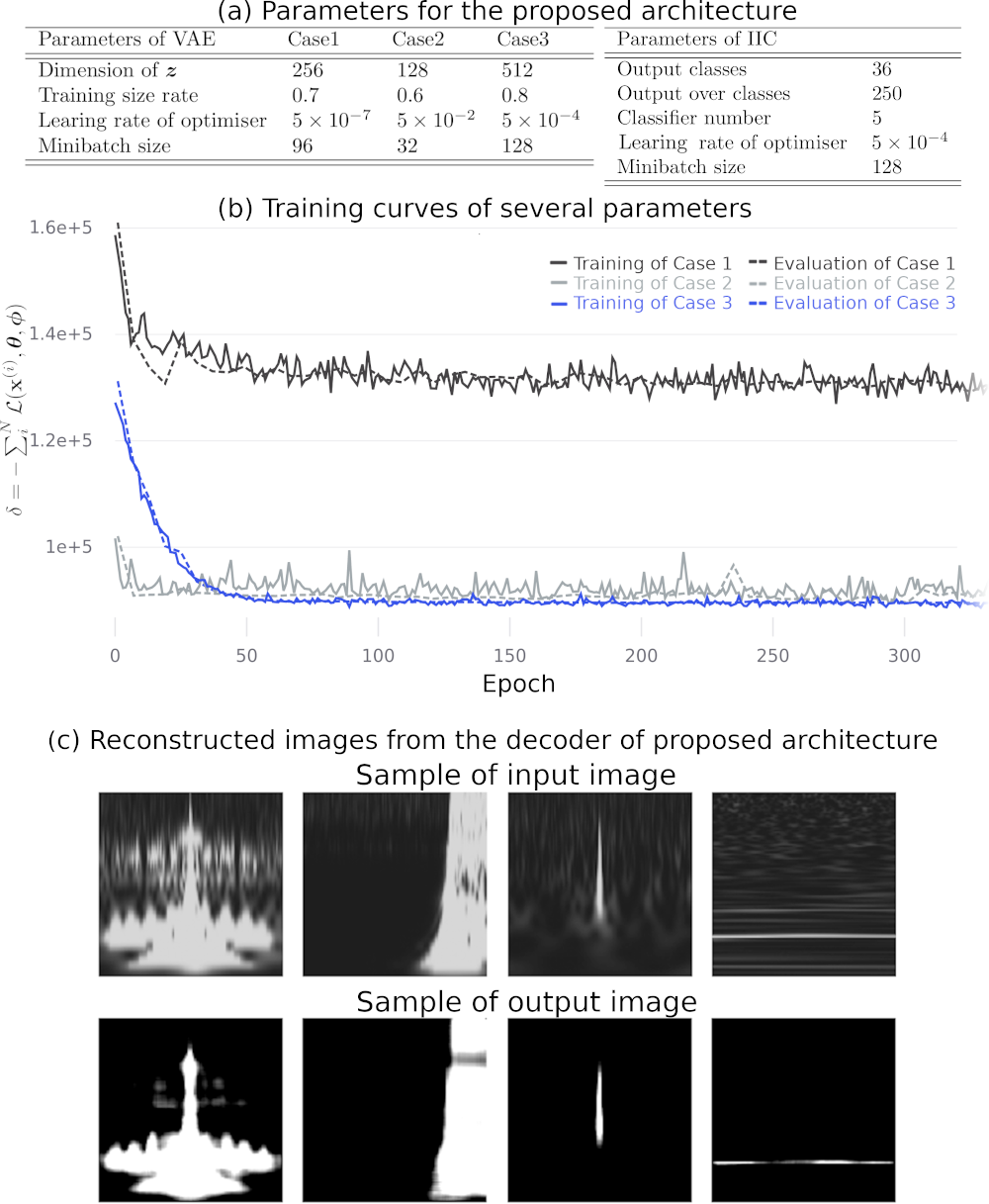}
  \caption{
  Left (a) Training parameters for the VAE of the proposed architecture.
  The dimension of $\bm{z}$ is the output number of the encoder. The training size rate is the ratio of the total number of data to the data size of the input at training. Regarding the architecture evaluation, the input size is set to $(1 - \mathrm{Training\, size\, rate})$. The learning rate is the initial learning rate, and the optimiser used is Adam~\cite{ref:Adam}. 
  Right (a) Training parameters for the IIC of the proposed architecture.
  The number of output classes is set to the number of classes to be classified.
  The classifier number is for multiple classifiers that are used to improve the performance of the classifier using spectral clustering.
  (b) Training curve during the training and evaluation of the VAE. The solid and dashed lines in the figure show the training objective $\delta \equiv -\sum_{i}^ {N}\vlb$ at the time of training and evaluation, respectively.
  (c) Reconstructed images generated by the decoder of the VAE at 100 epochs in Case 3.
  }
  \label{fig:learning_curve}
\end{figure}

\begin{figure}[t]
  \centering
  \includegraphics[scale=2]{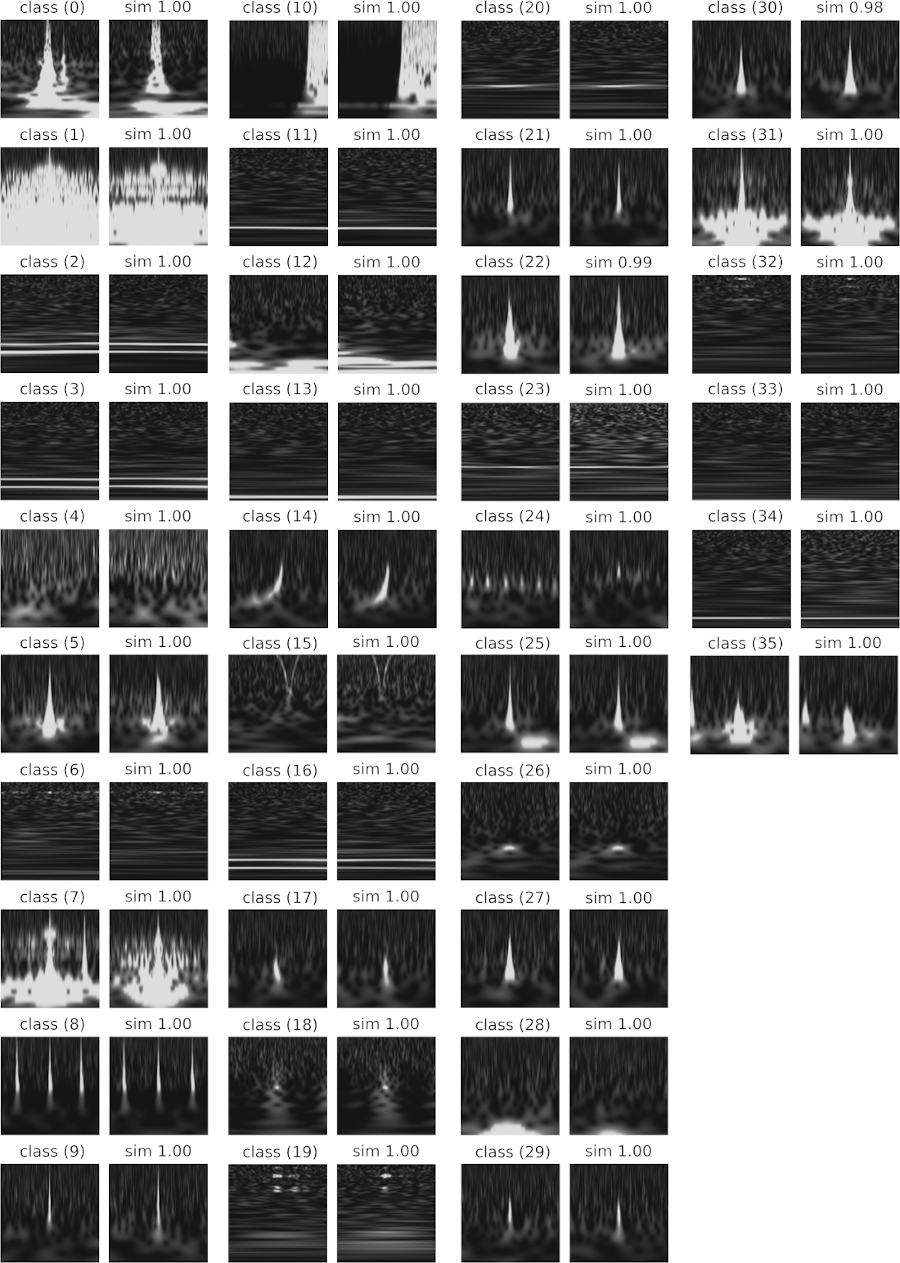}
  \caption{
    The representative and similar images in all the classes were classified using unsupervised learning. This representative image which is denoted by $i$ in the image is randomly selected from a class $i\in c = \{0,\dots, 35\}$, and its most similar image is to the right of a representative image in class ($i$). The cosine similarity to the representative image in class $i$ is shown at the top of the image.
    }
  \label{fig:all_class}
\end{figure}

\begin{figure}[t]
  \centering
  \includegraphics[scale=1.7]{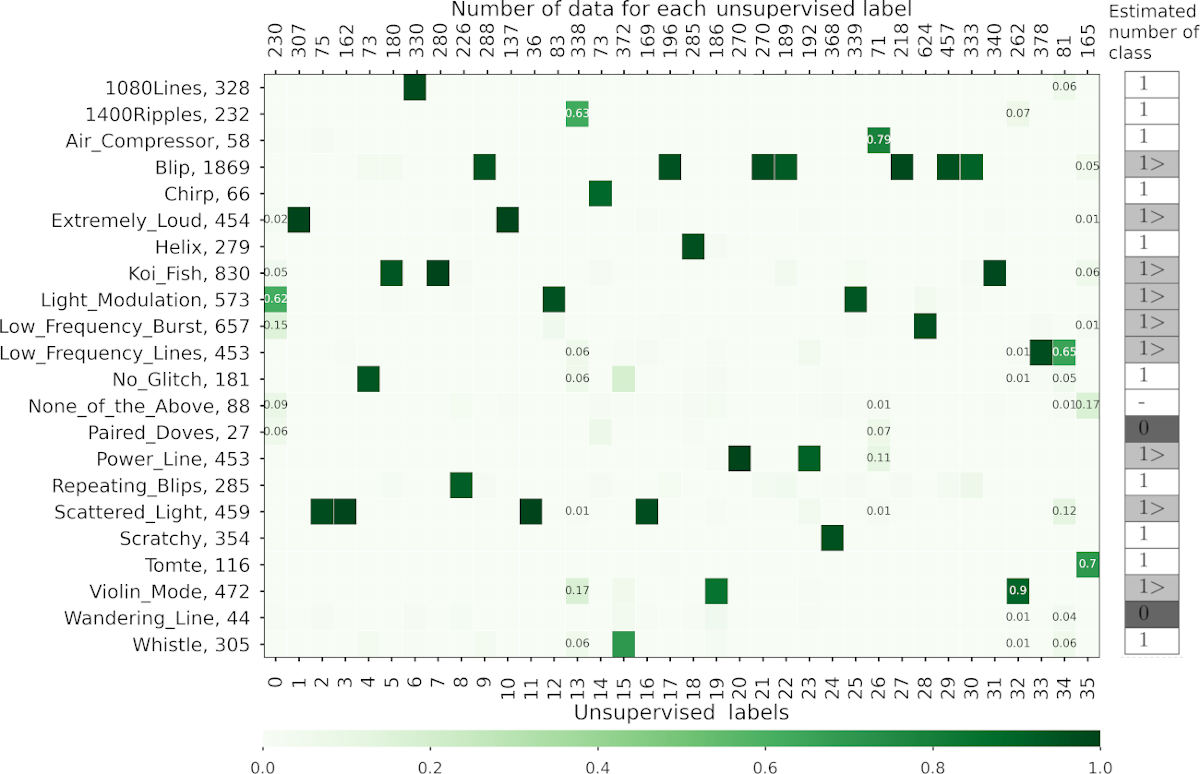}
  \caption{Confusion matrix of the classification results of the proposed architecture. The vertical axis of the confusion matrix represents the labels and number of data in the Gravity Spy dataset. The lower and upper horizontal axes denote the number of images classified into the unsupervised classes and the labels of the unsupervised classes, respectively. Each column of the confusion matrix is coloured using the ratio of the Gravity Spy-labelled images classified into the unsupervised class ($i$). In addition, the classes that are separated from the Gravity Spy labels on the confusion matrix, such as classes (0), (13), (26), (32), (35), and (36),  also show the ratio values in the matrix. The potential number of classes on Gravity Spy labels which are estimated by unsupervised learning are shown in the right column of the figure.  The notation ``1'' (in white cells) indicates that the number of classes labelled by the Gravity Spy matches the result of the unsupervised learning, and the inequality sign (in light grey cells) indicates that the class is separated into multiple classes in the unsupervised learning. The notation ``0'' (in dark grey cells) indicates an unclassified class in this training and dataset, and ``-'' notation indicates that they do not belong to any class of the unsupervised learning.
  }
  \label{fig:cm}
\end{figure}

\begin{figure}[t]
  \centering
  \includegraphics[scale=2.1]{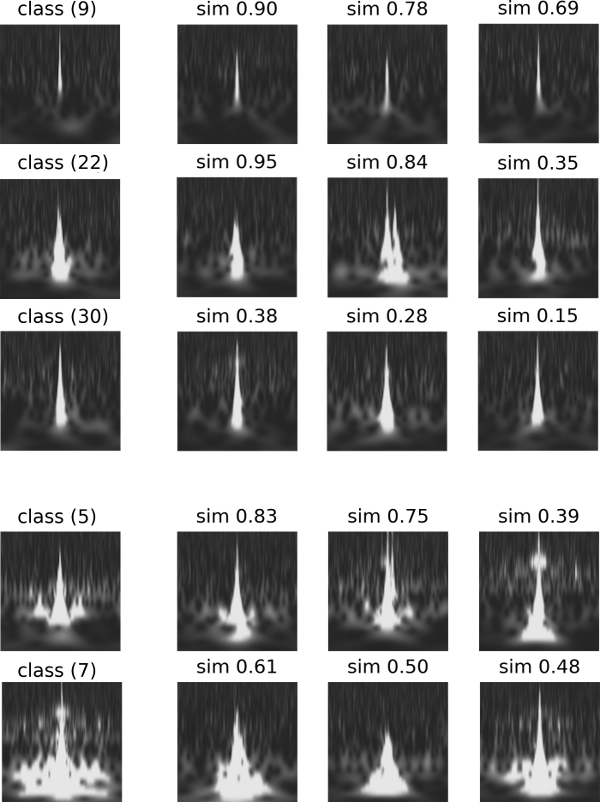}
  \caption{
    Representative images and images similar to unsupervised learning. Considering the figure, classes (9), (22), and (30) are separated from the` `Blip'' class, and classes (5) and (7) are separated from the``Koi\_Fish'' class. The representative images in the left column are sampled randomly from the images classified in class ($i$) using unsupervised learning. The similar images in the other columns are sorted in a descending order and are sampled randomly from the cosine similarity (a value at the top of an image), considering the representative image.
  }
  \label{fig:similar_sample}  
\end{figure}

\begin{figure}[t]
  \centering
  \includegraphics[scale=1.8]{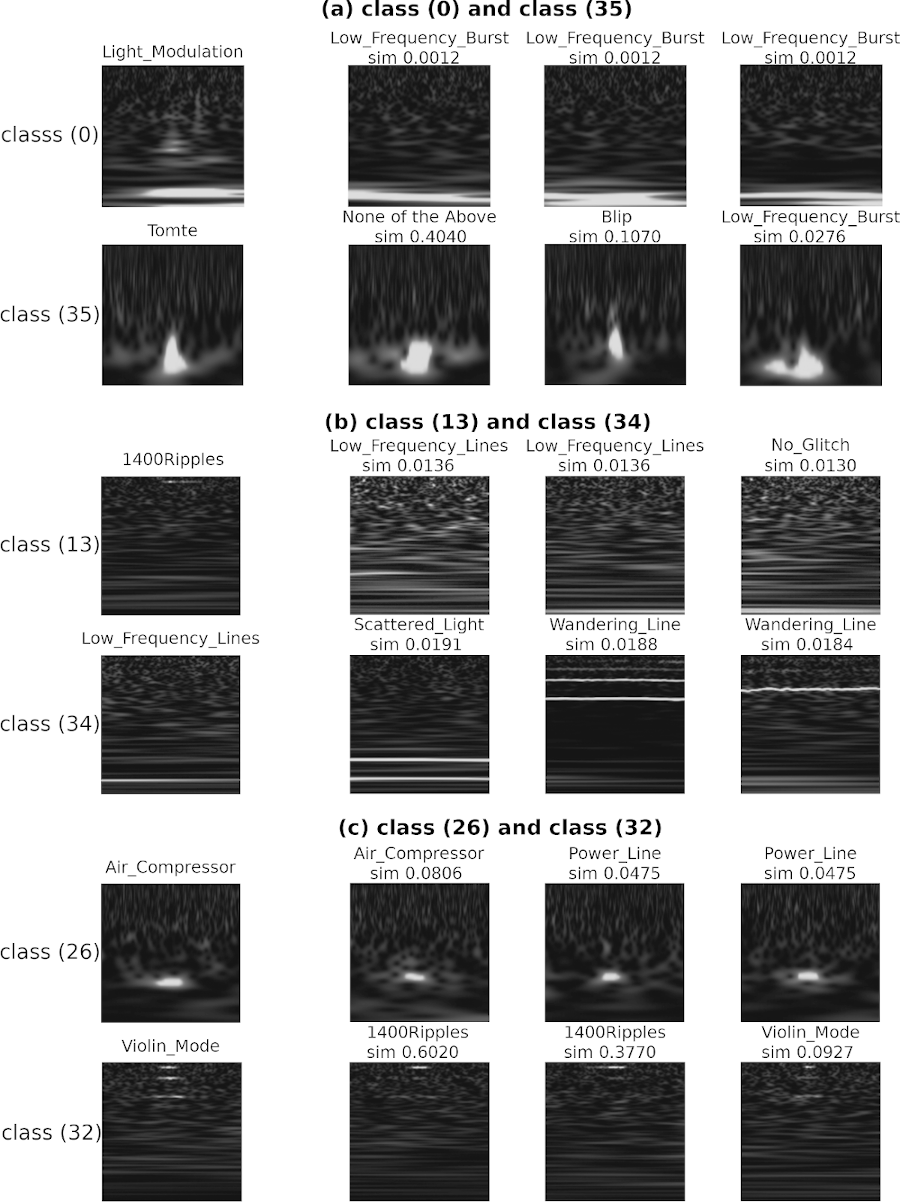}
  \caption{
  Examples of images in the classes with reduced accuracy in unsupervised learning. The major images in the left column are randomly sampled data from class $i$. The minor images in the other columns are sorted in an ascending order from the cosine similarity to its major image, indicating that they are sampled from the lowest similarity to the major one. The Gravity Spy label and the value of the cosine similarity are on top of the sampled image.
  }
  \label{fig:sample_worse}
\end{figure}

The Gravity Spy dataset~\cite{ref:Bahaadini2018}, which is the input dataset, is an image set of transient noise obtained from the LIGO O1~\cite{ref:Abbott2019}. Omicron software~\cite{ref:Robinet2020} searches for transient noise in time-series data, and Omega Scan~\cite{ref:Chatterji2004} software generates an image of the time--frequency spectrogram of each transient noise using Q-transformation~\cite{ref:Chatterji2004,ref:Brown1991}. Q-transformation is a method that estimates the frequency component of the time-series data by setting a window function on each time--frequency component, generating a 2D image of the time--frequency spectrogram. The spectrogram image of each transient noise in the Gravity Spy dataset has four time durations (0.5, 1.0, 2.0, and 4.0 s) at the centre, as shown in \figref{fig:data-classes}a. In addition, these transient noises are given 22 labels, which are related to cause as shown in \figref{fig:data-classes}b. For example, the images of 12 classes of transient noise are shown in \figref{fig:data-classes}c.

\subsection*{Pre-processing}
\begin{figure}[t]
  \centering
  \includegraphics[scale=2.0]{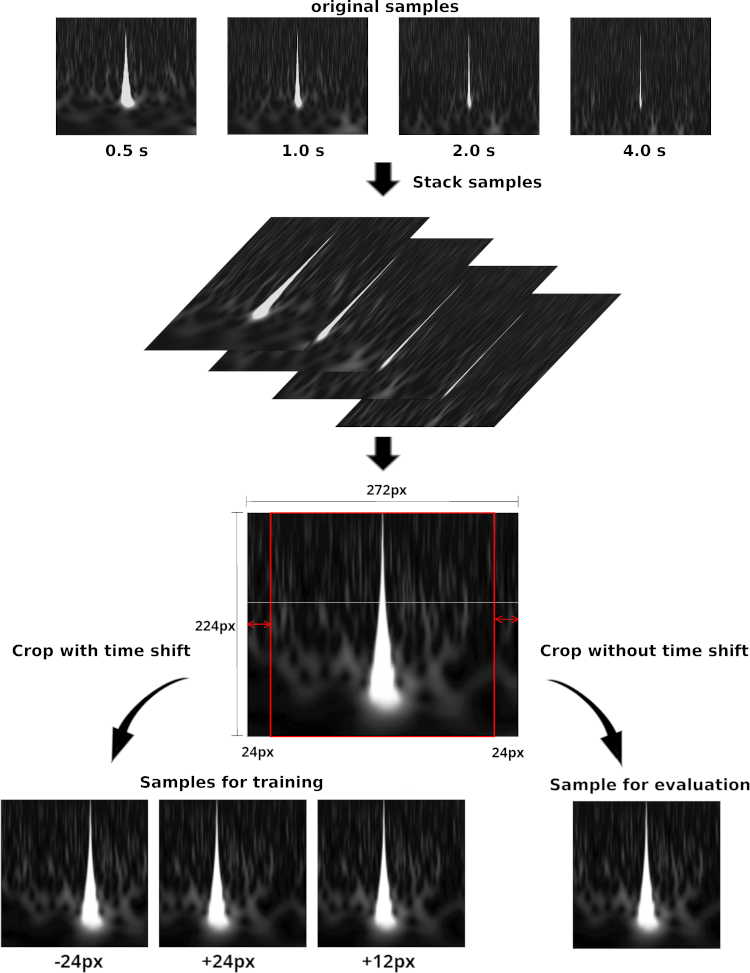}
  \caption{
  Overview of the input data pre-processing. The original samples are stacked in four time durations (0.5, 1.0, 2.0, and 4.0 s) to generate the data, where the width and height of each image are 224 px and 272 px, respectively, and the dimensions of the stacked data are 4, 224, and 272. Considering the training process of the proposed architecture, after a random time shift of image in the range 0--24 px, the dimensions of the data (4, 224, and 224) are cropped and the cropped data are used as training data. The data that are cropped without a time shift are used for the evaluation of the VAE and as the input image of the IIC. 
  }\label{fig:augmentation}
\end{figure}

The pre-processing applied to the Gravity Spy dataset for the training of our proposed architecture is shown in \figref{fig:augmentation}.

\begin{enumerate}
    \item For each transient noise, stack the images of the time--frequency spectrograms with the four time widths shown in \figref{fig:augmentation}, and use it as the input data for this transient noise.
    The resolution of the transient noise image for each time duration is 224 px $\times$ 272 px (frequency and temporal direction, respectively), and the dimensions of the stacked images are 4 $\times$ 224 $\times$ 272 px.
    \item Convert the stacked data into two types:
    \begin{description}
        \item[Input Image]: 
        Crop the left and right parts of the image equally such that the resulting image has dimensions of 4 $\times$ 224 px $\times$ 224 px
        \item[Perturbed Image]:
        Crop the left part of the image at the randomly time-shifted position in the range 0--24 px and also crop the right part of the image so that the resulting image has dimensions of 4 $\times$ 224 px $\times$ 224 px
    \end{description}
\end{enumerate}
Considering the characteristics of the time--frequency spectrogram, a small displacement in the time direction does not change its physical characteristics because this operation can be interpreted as a change in the event time. Therefore, the time-shifted images can be regarded as new events of transient noise, and it makes the architecture realise the classification of transient noise that does not depend on small displacements in the time direction. Conversely, a possible small displacement of the spectrogram in the frequency direction changes its physical characteristics.
Therefore, the frequency-shifted images fall into different classes to that of the original image in the classification.
Thus, the perturbation of transient noise is not applied in the frequency direction; nonetheless, they are applied only in the time direction.

In the training process of the proposed architecture, there is a random time shift of the image in the 0--24 px range used for the training data.
The data that were cropped without a time shift were used for the evaluation of the VAE and the input image of the IIC. 

\subsection*{Variational autoencoder}
In this study, the features of transient noise are obtained from their time--frequency 2D spectrogram image using VAE, one of the approaches for 
feature learning~\cite{ref:Zhong2016,ref:Bengio2013} using convolutional deep learning. Generally, feature learning is a method for acquiring features that are effective for the prediction and classification of data. It also has the ability to convert high-dimensional data to low-dimensional features.

Let the input dataset be $\mathcal{D} = \{ \x^{(1)}, \dots, \x^{(N)} | \x^{(i)} \in \mathbb{R}^D, i = 1, \cdots, N\}$ and the marginal likelihood for $\mathcal{D}$ be 
$p_{\bm{\theta}} (\x^{(1)}, \dots, \x^{(N)})$, where $D$ is the dimension number, $N$ is the number of the input data, and $\theta$ are parameters for the architecture.  The objective of the learning is to maximise the marginal likelihood. When the dataset $\mathcal{D}$ is \textit{independent and identically distributed}, the log marginal likelihood becomes $\sum_{i=1}^N\mathrm{ln}p_{\bm{\theta}}({\x^{(i)}})$.
Consider that the inference architecture $\qzlx$ (also known as \textit{encoder}) approximates $\qzlx \simeq \pzlx$, where $\bm{z} \in \mathbb{R}^J$ is a feature variable and $J < D$. Therefore, the log marginal likelihood $\lnp{\theta}{\x^{(i)}}$ can be expressed as
\begin{eqnarray}
  \lnp{\theta}{\x^{(i)}} = \ln 
                               \intz{p_{\bm{\theta}}(\x^{(i)}, \bm{z})}
                           \geq \intz{\qzlx \ln
                                  \left(
                            \frac{p_{\bm{\theta}}(\x^{(i)}, \bm{z})}{\qzlx}
                                  \right)}
                         \equiv \vlb.
\end{eqnarray}
The second inequality is obtained by the Jensen's inequality, and $\vlb$ is an objective function known as the \textit{lower bound}. Let a prior and a posterior distribution of $\bm{z}$ be a multivariate Gaussian distribution, indicating that $\pz = \mathcal{N}(\bm{z}|\bm{0}, \bm{I})$ and  $\qzlx = \mathcal{N}(\bm{z}|\bm{\mu_\phi}(\x^{(i)}),\bm{\Sigma_\phi}^{2}(\x^{(i)})\bm{I})$, where $\bm{\mu_\phi}(\cdot)$ and $\bm{\Sigma_\phi}(\cdot)$ are the outputs from an encoder and $\bm{I}$ is the identity matrix of dimension $J$. Let a posterior distribution of $\x$ be the multivariate Bernoulli distribution, $\pxlz = \mathrm{bern}({\x}^{(i)}|\bm{g}_{\bm{\theta}}(\bm{z}))$,
where $\bm{g}_{\bm{\theta}}(\cdot)$ are the outputs from the \textit{decoder}. Thus, the expression of the lower bound to be maximised is
\begin{equation}
  \vlb \simeq -D_{\mathrm{KL}}\left(
    \mathcal{N}(\bm{z}|\bm{\mu_\phi}(\x^{(i)}),\bm{\Sigma_\phi}^{2}(\x^{(i)})\bm{I})       ||
    \mathcal{N}(\bm{z}|\bm{0}, \bm{I})
  \right)
  + \frac{1}{L}\sum_{l=1}^L\ln \mathrm{bern}({\x}^{(i)}| \bm{z}^{(i,l)}),
  \label{eq:vlb}
\end{equation}
where $D_{\mathrm{KL}}[\cdot || \cdot]$ is the Kullback--Leibler divergence of two distributions and $\bm{z}^{(i,l)}$ is referred to as the \textit{reparameterisation trick},
such that $\bm{z}^{(i,l)} = \bm{g}_{\bm{\phi}}(\bm{\epsilon}^{(l)}, \x^{(i)}) = \bm{\mu}_{\bm{\phi}}(\x^{(i)}) + \bm{\epsilon}^{(l)}\odot
\bm{\Sigma}_{\bm{\phi}}(\x^{(i)})$, where $\bm\epsilon \sim \mathcal{N}(\bm0, \bm{I})$, and $\odot$ signifies the Hadamard product.

\subsection*{Classification using invariant information clustering}
A typical method for clustering is the $k$-means, which uses the Euclidean distances between data. Recently, several variants of the $k$-means have been developed (e.g. $k$-means++~\cite{ref:Arthur2007}, fuzzy $c$-means~\cite{ref:Bezdek1984}, and $x$-means~\cite{ref:Radwan2020}).
Regarding clustering in a high dimensional space, the variance of the distance between data becomes small owing to the ``curse of dimensionality''. Alternatively, IIC~\cite{ref:Ji2019}, which is a classification method, seems to be effective because it does not use the distances of the data for learning. In this study, transient noise is classified using IIC by maximising the mutual information. Let $\x \in \mathbb{R}^D$ be the input data, $\x^{\prime}$ be the perturbed data of $\x$, $C$ the number of output classes, and $\bm{\Phi}(\x) \in  \mathbb{R}^C$ be a classifier in which the output layer of the classifier uses the SoftMax activation function. Consider a pair of cluster assignments for two inputs, $\x$ and $\x^{\prime}$. Their conditional joint distributions and marginal distributions are 
$P_{ij} = \bm{\Phi}(\x^{(i)}) \cdot \bm{\Phi}(\x^{(j)\prime})^{\mathrm{T}}$ and $P_{i} = \sum_j\bm{\Phi}(\x^{(i)}) \cdot \bm{\Phi}(\x^{(j)\prime})^{\mathrm{T}}$,
respectively, where the superscript $\mathrm{T}$ denotes the transpose. The objective for the maximisation of the mutual information is expressed as
\begin{equation}
  \max_{\bm{\Phi}}I(\bm{\Phi}(\x), \bm{\Phi}(\x^\prime))
  = \sum_i^C\sum_j^C P_{ij}\ln \frac{P_{ij}}{P_i P_j}.
  \label{eq:maximum_mutual_info}
\end{equation}
To improve the performance of the classifier, auxiliary over-clustering~\cite{ref:Ji2019} is also used when calculating the mutual information.
This over-clustering formula is the same as \eqref{eq:maximum_mutual_info}, except for $\bm{\Phi}(\x) \in \mathbb{R}^W$, where $C < W$.

\subsection*{Proposed architecture}
\begin{figure}[t]
  \centering
  \includegraphics[scale=1.5]{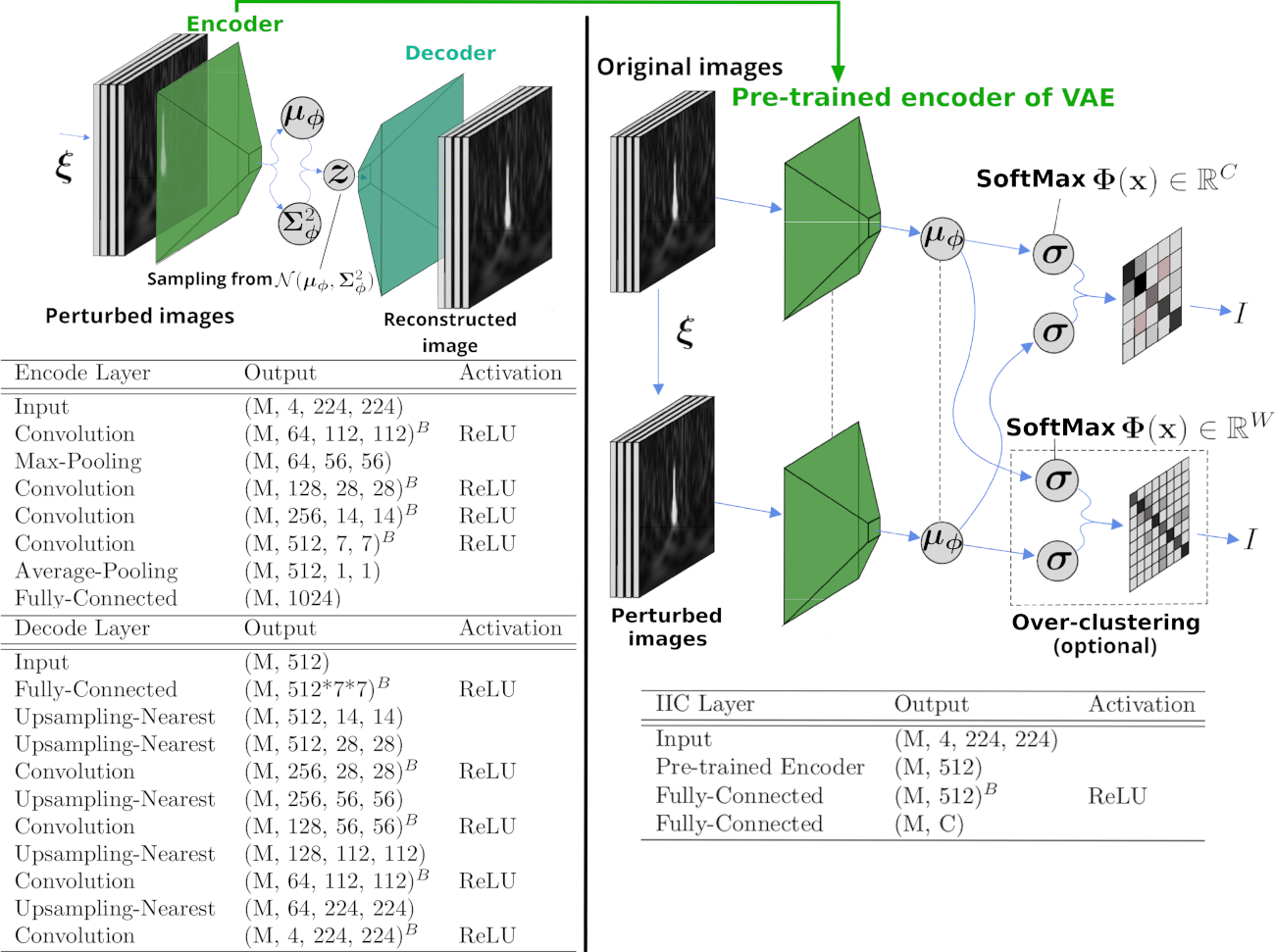}
  \caption{
    Proposed architecture for the classification of transient noise. The tables show the details of the architectures of neural networks. $^B$ denotes batch-normalise to an object, and $M$ denotes the mini-batch size. Left: Schematic architecture of the VAE for feature learning. The VAE trains neural networks to maximise the lower bound in \eqref{eq:vlb}. The input to the VAE is a perturbed image $\x^{\prime}$ of the time--frequency spectrogram of the transient noise. This pre-process allows the encoder to learn features that do not depend on the perturbation. At the output layer of the encoder, the average and variance of the feature variable $\bm{z}$ are output from the same network and separated into the dimensions (M, 512). Subsequently, the feature variables $\bm{z}$ are constructed using the \textit{reparameterisation trick}. The decoder uses $\bm{z}$ to generate a reconstructed image that is close to the input image. Right: Schematic architecture of the IIC for classification. The IIC trains neural networks to maximise the mutual information between the input data and its perturbed data. The inputs to the pre-trained encoders of the VAE are the original and perturbed images, respectively. Both encoders have the same architecture, and the dashed lines indicate the sharing weights of the neural networks in the figure. The IIC classifies transient noise using the SoftMax activation function at the output layer from the feature, which is the output of the pre-trained encoder. $C$ is the estimated number of classes of the transient noise, and $W$ is the number of classes used in the over-clustering.
  }
  \label{fig:architecture}  
\end{figure}

We propose the unsupervised classification architecture shown in \figref{fig:architecture}. It is a deep learning architecture that trains time--frequency 2D spectrogram images of transient noise. Considering the proposed architecture, the feature variables of the input image $\x$ and its perturbation image $\x^\prime = \bm\xi(\x)$ are extracted by a pre-trained encoder of the VAE. The perturbation $\bm\xi$ is a transformation that does not change the information required for the classification (see \secref{Pre-processing}). Subsequently, the IIC learns to maximise the mutual information $I(\bm{\Phi}(\bm{z}), \bm{\Phi}(\bm{z}^{\prime}))$, which is composed of a pair of feature variables $(\bm{z} = \bm{\mu}_{\bm\phi}(\x), \bm{z}^{\prime} = \bm{\mu}_{\bm\phi}(\x^{\prime}))$.

The clustering of the IIC depends on the initial values of the neural networks in which the values are randomly provided. Thus, the classification results from each classifier varies slightly. Regarding the unsupervised learning, it is difficult to apply an ensemble average for each classification result to solve the dependencies of the initial values because the classified labels are random at each time. In this study, spectral clustering~\cite{ref:VonLuxburg2007} was applied to compress the multiple results of classification into one result. The procedure is as follows:
\begin{enumerate}
    \item Let $D$, $K$, and $C$ be the number of datasets, number of classifiers, and estimated number of classes, respectively. Create a \textit{hypermatrix } $H$ whose dimension is $(D, K \times C)$ from each classifier result.
    \item Considering $h^{(i)}$ as a row vector for each data of $H$, calculate an affinity matrix using the Gaussian kernel, which is given by $\mathrm{exp}(-\|h^{(i)}-h^{(j)}\| )$.
    \item Compute the spectral clustering to the affinity matrix. Consequently, we have labels for each dataset whose dimension is $(D, C)$.
\end{enumerate}




\section*{Acknowledgements}
We are grateful to the members of the Gravity Spy project for enlightening discussions. 
This study was supported in part by the Inter-University Research Program of Institute for Cosmic Ray Research, University of Tokyo, Japan. 
It was also supported in part by the Japan Society for the Promotion of Science (JSPS) Grants-in-Aid for Scientific Research on Innovative Areas, Grant No.~24103005 [JP17H06358, JP17H06361, and JP20H04731], by JSPS Core-to-Core Program A, Advanced Research Networks, and by JSPS KAKENHI [Grant No.~19H0190 (YI and HT) and Nos.~19K14636 and 21H05599 (Y.~Shikano)],  and by JST, PRESTO [Grant No. JPMJPR20M4 (Y.~Shikano) ].

\section*{Author contributions}
Y.Sakai, G.U., and H.T. conceptualised the study; Y.Sakai, G.U., Y.Sikano, T.U., and H.T. framed the methodology; Y.Sakai, G.U., and H.T. concentrated on the software development and analysis; Y.I., P.J., K.K., C.K., K.T.N., S.O., Y.Shikano, T.U., T.W., T.Yamamoto, and T.Yokozawa performed the validation and investigation; K.K., C.K., S.O., T.W., T.Yamamoto, and T.Yokozawa made some critical and technical suggestions; Y.Sakai, G.U., and H.T. prepared the initial draft of the manuscript; all the authors, mainly Y.Sakai, Y.I., Y.Shikano, H.T., and T.U. wrote, reviewed, and edited the manuscript; H.T. supervised the study; H.T. and T.U. administered the project. All authors have read and agreed to the published version of the manuscript.

\section*{Data availability}
All the results are reported for public data of Gravity Spy project.
Data on the results of unsupervised learning are available upon request from Y.~Sakai and HT.

\section*{Code availability}
All the codes developed in this study are available upon request from Y.~Sakai and HT.

\section*{Competing interests}
The authors declare no competing interests.

\section*{Additional information}
\textbf{Supplementary Information} The online version contains supplementary material available at \href{https://doi.org/10.1038/s41598-022-13329-4}{https://doi.org/10.1038/s41598-022-13329-4}.\\
\textbf{Correspondence} and requests for materials should be addressed to Y.~Sakai (g2191402@tcu.ac.jp) and HT (hirotaka@tcu.ac.jp).\\
\textbf{Reprints and permissions information} is available at \href{https://www.nature.com/nature-portfolio/reprints-and-permissions}{www.nature.com/reprints}.\\
\textbf{Publisher's note} Springer Nature remains neutral with regard to jurisdictional claims in the published maps and institutional affiliations.

\appendix
\renewcommand{\thefigure}{S\arabic{figure}}
\setcounter{figure}{0}
\section*{Feature Visualization of Transient Noise using t-SNE}
Generally, the feature space for the input data is high-dimensional, and it can be expressed visually by reducing it to a low-dimensional space. One of the nonlinear dimensionality reductions~\cite{ref:Roweis2000} is the t-distributed stochastic neighbour embedding~\cite{ref:Maaten2008} (t-SNE). 
The t-SNE algorithm embeds the data in high-dimensional space as a t-SNE component in low-dimensional space, and this embedding is effective for visualising the data in a high-dimensional space.

\begin{figure}[t]
  \centering
  \includegraphics[scale=2.0]{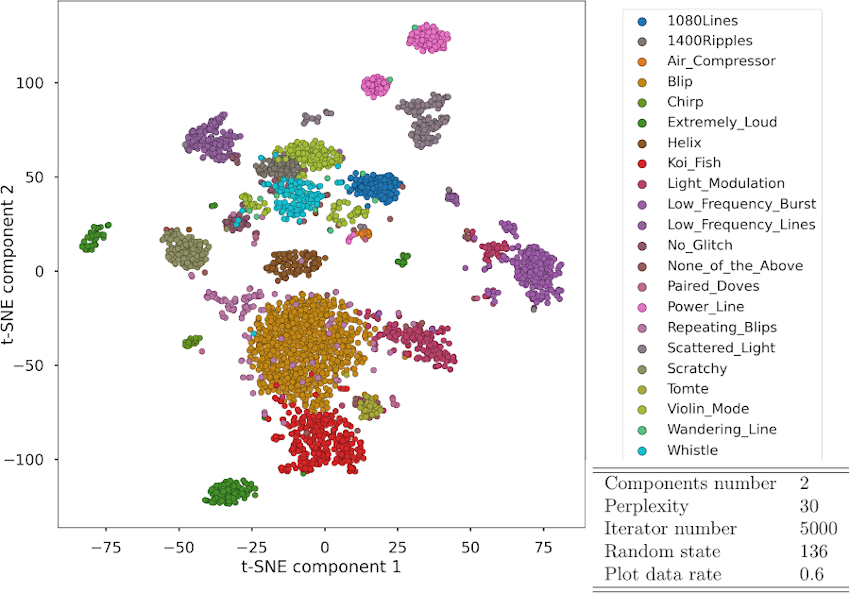}
  \caption{Left: Visualisation of the feature of transient noise. The visualisation is computed by a composition mapping $(h\circ f)$, where $f$ is the pre-trained encoder of the VAE, which is a mapping from the input space $\mathbb{R}^{224 \times 224 \times 4}$ to the feature space $\mathbb{R}^{512}$. $h$ is t-SNE, which is an embedding from the feature space $\mathbb{R}^{512}$ to the 2D plane. By using the Gravity Spy labels for the input data, the embedding data have labels that are useful for visualisation on the plane. Top right: List of colours corresponding to the Gravity Spy labels.  Bottom right: embedding parameters used for the t-SNE. The number of components indicates the number of dimensions after the embedding. Perplexity is a characteristic that is related to the number of nearest neighbours, and the iterator number refers to the maximum number of optimisations for exploring the nearest neighbours. The plotted data rate is the ratio of the number of plots to the entire dataset.}
  \label{fig:tsne}
\end{figure}

In this study, a composition mapping $h\circ f: \mathbb{R}^{224 \times 224 \times 4} \to \mathbb{R}^2$ is used to evaluate the features of the transient noise,
where $f: \mathbb{R}^{224 \times 224 \times 4} \to \mathbb{R}^{512}$ is a mapping from the input space to the feature space using a pre-trained encoder of the VAE
and $h: \mathbb{R}^{512} \to \mathbb{R}^2$ is an embedding from the feature space to a 2D plane using t-SNE. The feature visualisation of transient noise is shown at the left of \figref{fig:tsne}, and the parameters used for the embedding are shown at the bottom right of \figref{fig:tsne}.

We investigated how the Gravity Spy dataset is clustered in the feature space by visualisation using t-SNE. Considering the data on the Gravity Spy labels of ``1080Lines'', ``Chirp'', ``Helix'', ``Scratchy'', and ``Tomte'', it can be observed that they are formed closely on the feature visualisation. Therefore, the features of the transient noise by unsupervised learning are consistent with the Gravity Spy classes. Regarding the data of ``Power\_Line'', ``Extremely\_Loud'', ``Scattered\_Light'', and ``Violin\_Mode'',
these distributions are separated on feature visualisation. Similar results for the separation of these data have been obtained in the unsupervised classification (see Figure 4). Considering the data of ``None\_of\_the\_Above'', which are not classified as a unique class on the confusion matrix shown in Figure 4, it can be confirmed that they are not formed closely, even on the feature visualisation.  Therefore, consistent results on the above labels/classes were obtained between the feature visualisation and the result of unsupervised classification.

On the contrary, considering the data of ``Blip'', ``Koi\_Fish'' are separated into subclasses on the confusion matrix shown in Figure 4, whereas ``Blip'' and ``Koi\_Fish'' are formed as one distribution, respectively, on feature visualisation in \figref{fig:tsne}. This formation seems to have resulted in the superposition of small classes by embedding the data into a 2D plane. Because the distribution of ``Blip'' and ``Koi\_Fish'' must be separated at higher dimensions, the result of the unsupervised learning shown in Figure 4 implies that ``Blip'' and ``Koi\_Fish'' have subclasses.

\end{document}